\documentclass[]{spie}  
\usepackage[]{graphicx}

\title{Near-Infrared InGaAs Detectors for Background-limited Imaging and Photometry} 

\author{Peter W. Sullivan\supit{a}, Bryce Croll\supit{a,}\supit{b}, and Robert A. Simcoe\supit{a}
\skiplinehalf
\supit{a}MIT-Kavli Institute for Astrophysics, 77 Massachusetts Ave., Cambridge, MA, USA; \\
\supit{b}NASA Carl Sagan Fellow
}

\authorinfo{Send correspondence to: pwsully@mit.edu}

\begin{document} 
  \maketitle 

\begin{abstract}
Originally designed for night-vision equipment, InGaAs detectors are beginning to achieve background-limited performance in broadband imaging from the ground. The lower cost of these detectors can enable multi-band instruments, arrays of small telescopes, and large focal planes that would be uneconomical with high-performance HgCdTe detectors. We developed a camera to operate the FLIR AP1121 sensor using deep thermoelectric cooling and up-the-ramp sampling to minimize noise. We measured a dark current of 163$~e$- s$^{-1}$ pix$^{-1}$, a read noise of 87$~e$- up-the-ramp, and a well depth of 80k~$e$-. Laboratory photometric testing achieved a stability of 230 ppm hr$^{-1/2}$, which would be required for detecting exoplanet transits. InGaAs detectors are also applicable to other branches of near-infrared time-domain astronomy, ranging from brown dwarf weather to gravitational wave follow-up.

\end{abstract}



\section{INTRODUCTION}
\label{sec:intro}  
Off-the-shelf imaging detectors made from InGaAs have the potential to greatly reduce the cost of near-infrared, ground-based instrumentation. The quality of domestically-produced InGaAs has dramatically improved in recent years, and focal planes are becoming available in formats suitable for imaging and photometry. InGaAs-based instruments may be particularly appropriate for small telescopes where high-performance HgCdTe detectors are too costly; they could also be used in large mosaics or arrayed instruments where the price per pixel drives the cost.

As a direct-bandgap semiconductor, InGaAs has high quantum efficiency up to a cut-off wavelength of 1.7 $\mu$m, which is set by lattice-matching to the substrate. This cut-off allows its use across the $Y$, $J$, and most of the $H$ bands while having less sensitivity to thermal emission from the telescope and the sky than 2.5 $\mu$m HgCdTe detectors. In order to achieve background-limited imaging performance, the detector needs to be cooled to the point where the dark current per pixel is less than the sky surface brightness, and the read noise must be lower than the Poisson noise from the dark current and the sky. 

If such temperatures can be reached with thermoelectric rather than cryogenic cooling, then far less expensive instruments can be designed. Furthermore, if commercial readout integrated circuits (ROICs) can provide low read noise, then off-the-shelf sensor packages can be used. Such ROICs generally use capacitive transimpedance amplifier (CTIA) pixel architectures rather than a source follower, so their read noise is higher. However, the CTIA and snapshot integration on commercial ROICs can potentially reduce the nonlinearity of a source follower as well as the systematic errors of a rolling electronic shutter.

Beginning two years ago, we characterized the FLIR AP640C detector for astronomical applications \cite{sullivan13}. While the read noise of this device was acceptable (53$~e$- in sample-up-the-ramp), readout glow created a dark current floor of 840$~e$- s$^{-1}$ pix$^{-1}$ that limited the utility of the detector. In 2013, we began testing a new detector from FLIR, the AP1121, that offered reduced background. The detector is VGA-format with 640$\times$512 active pixels, and its performance is representative of upcoming 1.9K$\times$1K arrays. We will present and discuss the results of the AP1121 testing in this paper.

\section{IMPLEMENTATION}
Commercially-available InGaAs sensors are often optimized for high-cadence video, but we have developed our own camera with analog-to-digital electronics and firmware optimized for low noise and high stability. In particular, we use regular clocking patterns to keep the detector in thermal equilibrium, precise temperature control with on-chip sensing, and 16-bit analog-to-digital conversion. We employ closed-loop water cooling to draw heat from the detector's thermoelectric cooler. Linear power supplies drive both the camera electronics and the thermoelectric cooler. An Opal-Kelly FPGA module provides a USB interface to a host computer.

Under normal operations, the detector is read nondestructively with a frame rate of 3 Hz, so exposures lasting tens of seconds will have up to 100 frames for processing up-the-ramp in order to reduce read noise (see Section \ref{sec:rn}). The AP1121 supports a variety of gain settings, but we characterize the detector in just one configuration here.

\section{NOISE MEASUREMENTS}
\label{sec:noise} 
In order to determine the gain of the detector, we used the photon transfer method \cite{janesick} to compare the variance against the signal in a series of flat-field images. We obtained a set of 100 flat-field images to measure the temporal variance and median signal at six exposure levels ranging from dark to half-full well. We calculated the gain on a pixel-by-pixel basis and found the median value to be 1.7 $e$-/ADU. We then used this data to calibrate measurements of the dark current and read noise (below).

\subsection{Dark Current}
\label{sec:dc}  
Our previous experience with the AP640C detector had shown that dark current from ROIC glow limited its performance. The AP1121 detector has improved masking between the readout and the photodiode array to reduce this source of background. Furthermore, the AP1121 has a smaller pixel size of 15 $\mu$m versus 25 $\mu$m, and we are able to operate the detector at -$50^{\circ}$C rather than -$40^{\circ}$C.

\begin{figure}
\begin{center}
\begin{tabular}{c}
\includegraphics[width=17.5cm]{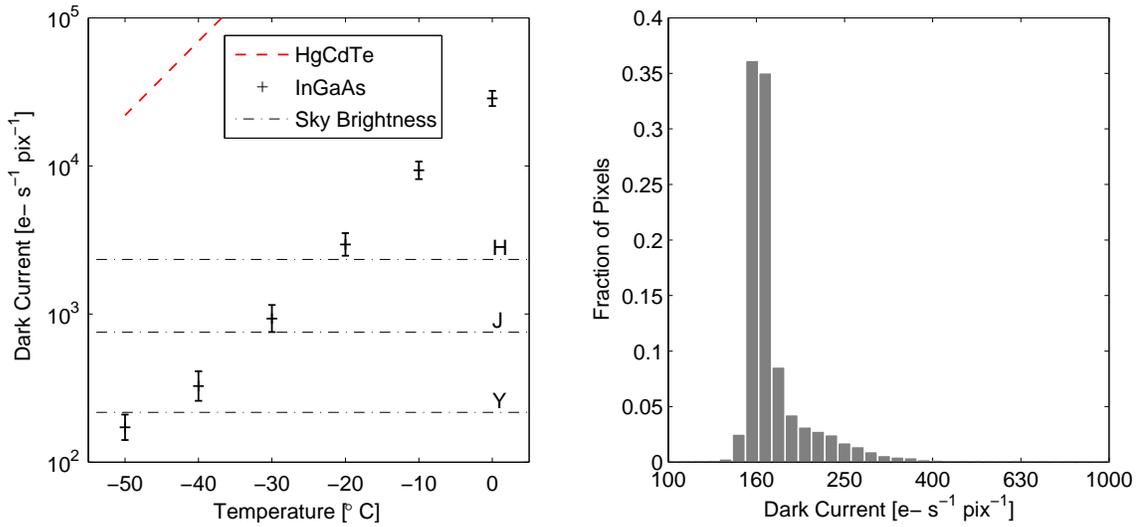}
\end{tabular}
\end{center}
\caption[example] 
{ \label{fig:dc} 
\textit{Left:} Thermoelectric cooling can reduce the dark current of InGaAs detectors below the sky surface brightness, which is plotted for the $Y$, $J$, and $H$ bands. We have assumed 1.0 m telescope with 1'' pixels. The dark current of HgCdTe \cite{beletic08} for the same cutoff wavelength and scaled to the same pixel size (red dashed line) is much higher, so cryogenic cooling is required to reach the same level of dark current. \textit{Right:} The distribution of dark current per pixel at -$50^{\circ}$C has a median value of 163$~e$- s$^{-1}$ pix$^{-1}$; 87$\%$ of pixels have dark current below 200$~e$- s$^{-1}$ pix$^{-1}$.}
\end{figure}

We find that the dark current scales exponentially temperature through -$50^{\circ}$C, halving with every 7$^{\circ}$C of cooling as expected. The median dark current at -$50^{\circ}$C is 163$~e$- s$^{-1}$ pix$^{-1}$, and 87$\%$ of pixels have dark current below 200$~e$- s$^{-1}$ pix$^{-1}$. Figure \ref{fig:dc} shows the temperature scaling and dark current distribution.

For comparison to the sky surface brightness, we use the broadband sky measurements from the FourStar camera\footnote{http://instrumentation.obs.carnegiescience.edu/FourStar/calibration.html} on the 6.5 Magellan Baade telescope. The scaling from a 6.5 m telescope with 0.159'' pixels to a 1.0 m telescope with 1.0'' pixels is nearly unity. As Figure \ref{fig:dc} shows, the dark current is well below the sky brightness in the $J$ and $H$ bands for a 1.0 m telescope. However, requiring sky-limited observations in the $Y$ band give a more strict constraint on the dark current.  The $Y$-band sky scales to 220$~e$- s$^{-1}$ arcsec$^{-2}$, so sky noise should dominate dark current noise for telescopes larger than 1.0 m. A larger telescope, or deeper cooling, is required if the pixels have a smaller projected area on the sky.

\begin{figure}
\begin{center}
\begin{tabular}{c}
\includegraphics[width=10cm]{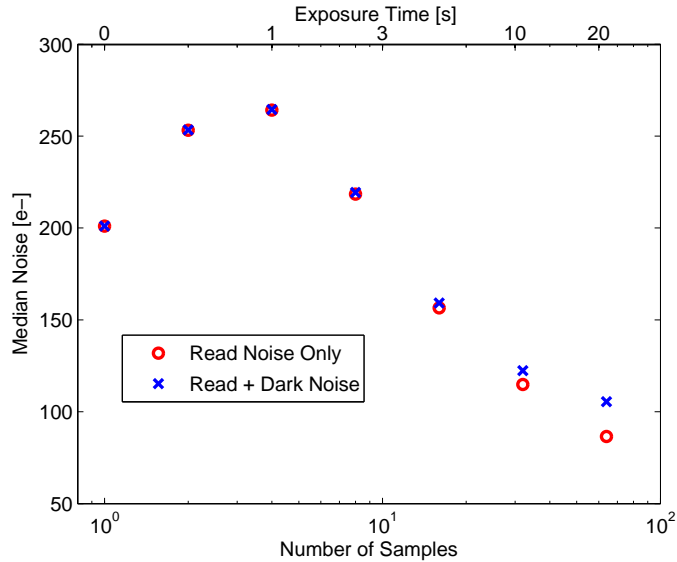}
\end{tabular}
\end{center}
\caption[example] 
{ \label{fig:rn} 
Recording multiple non-destructive samples is needed to overcome the read noise of a single exposure. As the exposure time lengthens from left (0 s) to right (21 s), sample-up-the-ramp processing can reduce the single-read noise by more than a factor of two (from 200$~e$- to to 87$~e$-). Since the noise was calculated on dark frames, subtracting the dark noise is necessary to measure the read noise of the longer exposures. The values here were measured at the telescope; the noise is $\sim10\%$ lower in the laboratory.}
\end{figure} 

\subsection{Read Noise}
\label{sec:rn}  
With the earlier APS640C detector, we found that we could substantially reduce the effective read noise by nondestructively sampling the detector up-the-ramp. We employ a similar approach with the AP1121, whose read noise can be reduced from 200$~e$- in single-sample reads to 87$~e$- in ramps of 64 samples (Figure \ref{fig:rn}). The read noise will nonetheless dominate the combined Poisson noise from the sky and dark current for exposures lasting less than 20 s on a 1.0 m telescope. We are still investigating other readout modes to optimize the read noise performance of the AP1121.

\section{DETECTOR RESPONSE}
\label{sec:resp}

\subsection{Nonlinearity}
\label{sec:lin} 
\begin{figure}
\begin{center}
\begin{tabular}{c}
\includegraphics[width=10cm]{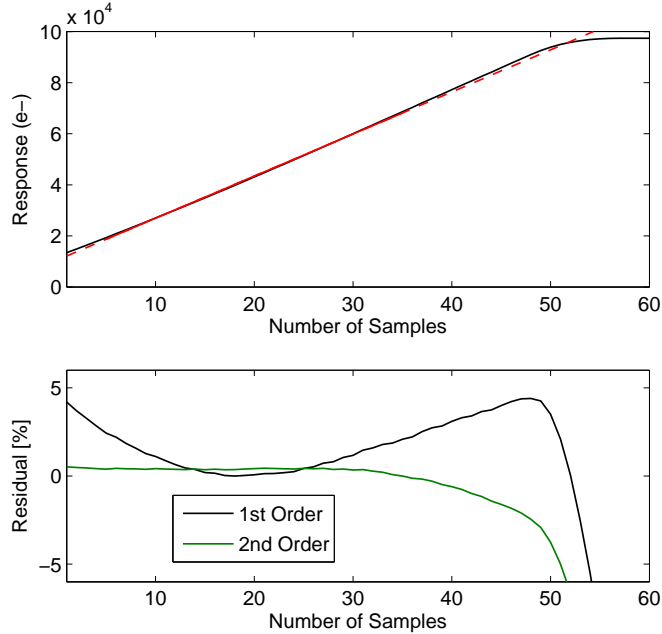}
\end{tabular}
\end{center}
\caption[example] 
{ \label{fig:lin} 
\textit{Top panel:} The response of the detector to a constant flux, plotted in black, shows a small degree of nonlinearity. A linear function is fitted to the response over the central region indicated by the solid red line. In order to quantify the nonlinearity outside of this region, the model is extended across all of the samples (red dashed line). \textit{Bottom panel:}  Subtracting the linear fit to the detector response shows the relative nonlinearity (black). Subtracting a second-order fit to the data shows the higher-order residuals (green).}
\end{figure}

The AP1121 exhibits a small degree of nonlinearity, which we quantified by obtaining several long series of up-the-ramp samples with illuminating the detector with a fixed flat-field source. The stack of these ramps is shown in Figure \ref{fig:lin}. We assume that equal numbers of photons are collected during the sampling intervals, so any nonlinearity in the ramp is due to the detector. After subtracting a best-fit line to the median ramp, we find that detector non-linearity is approximately 5$\%$. Removing a second-order polynomial reduces the residuals to the 0.5$\%$ level, showing that the nonlinearity can be easily corrected. The full well capacity for this mode, which we define to be the number of electrons collected before the nonlinearity turns downwards, is approximately 80k$~e$-. 

\subsection{Persistence}
\label{sec:persis}  
\begin{figure}
\begin{center}
\begin{tabular}{c}
\includegraphics[width=10cm]{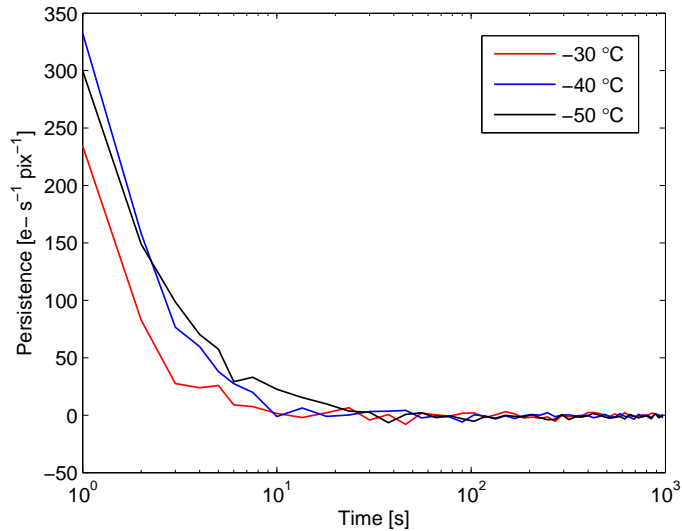}
\end{tabular}
\end{center}
\caption[example] 
{ \label{fig:persis} 
Persistence measured immediately after shuttering a bright flat-field source. No persistence is detected after 20 s, which is comparable to the time required to slew a telescope to a new target. Modestly bright targets should not persist in dithered images.}
\end{figure}

Next, we measured the persistence (latent image) by measuring the excess dark current of the detector immediately after being exposed to a high flux level. We exposed the detector to the high flux level of 100k$~e$- s$^{-1}$ for approximately one hour before removing the flux with a shutter. A series of 0.3 s exposures in CDS mode were then taken for the next 1000s. The timeseries is shown in Figure \ref{fig:persis}.

At our nominal operating temperature of -$50^{\circ}$C, the detector persistence falls below the dark current level ($\sim$200 $e$- s$^{-1}$ pix$^{-1}$) within 2 s. Within 20 s, no persistence is detectable. We repeated the experiment at -$40^{\circ}$C and -$30^{\circ}$C and found that the persistence is reduced at these higher temperatures. This is expected for persistence due to trapping sites; at lower temperatures, photoelectrons can be trapped for a longer period of time.

The detector characterized here did not have it substrate removed, so it is possible that the persistence originates in the substrate and not the InGaAs. However, we conducted a similar test with an earlier device with its substrate removed and found no difference in the persistence. While there are other advantages to removing the substrate (including blue-end response and cosmic ray reduction), the persistence is unaffected since it appears to originate in the InGaAs itself.

\section{PHOTOMETRIC PERFORMANCE}

\subsection{Laboratory Photometry}
\label{sec:lab}  
We characterized the differential photometric performance of the AP1121 using the same setup for the AP640C \cite{sullivan13}, which was inspired by a similar test carried out for a HAWAII-2RG detector.\cite{clan12} A lenslet array is used to re-image a pinhole into a 9$\times$7 grid of simulated stars. We obtained a long time series of 5 s exposures as well as flat-field and dark frames. The contributions to the noise budget are summarized in Table \ref{tab:noise}.

After applying the dark and flat-field corrections, we performed aperture photometry on each of the 63 stars generated with the lenslet array. The apertures had a median size of 14 pixels and received 2.35$\times 10^5~e$- per exposure. We also measured the flux centroid of each star; the centroids moved by $\sim$0.015 pixels during the data acquisition. 

Over shorter (minute-long) timescales, the observed noise shown in Table \ref{tab:noise} is 16$\%$ \textit{lower} than the expected value. We suspect this is due to the subtraction of common-mode noise in the aperture photometry; since the read noise was calculated on flat fields, we could not easily subtract common-mode noise. 

Before measuring the long-term photometric stability of each star, we de-trended four quantities from the photometric timeseries using a robust linear regression: the median $x$ and $y$ centroid location, the median flux, and the median dark level that was subtracted from each frame. These signals should induce systematic errors from intra-pixel sensitivity variations, changes in the lamp output, and over- or under-subtraction of the dark value. For a given star, we were careful to use the median centroids and fluxes of the \textit{other} stars when de-trending. 

\begin{figure}
\begin{center}
\begin{tabular}{c}
\includegraphics[width=10cm]{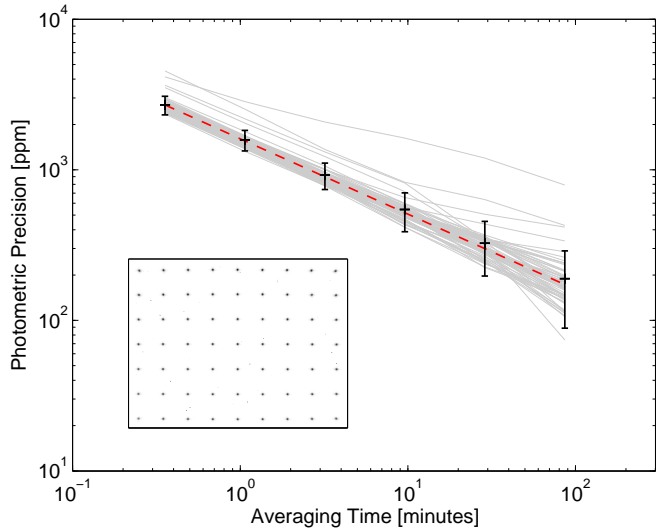}
\end{tabular}
\end{center}
\caption[example] 
{ \label{fig:lab} 
We measured the AP1121 photometric performance on a grid of 63 stars generated with a lenslet array (\textit{inset}). The relative noise for each star is plotted against co-averaging time in gray lines; the mean and standard deviation across the 63 stars is plotted with error bars. The red, dashed line shows a precision of 206 ppm hr$^{-1/2}$, which is the minimum expected for uncorrelated noise scaled to the leftmost data point. We achieved an actual precision of 230 ppm hr$^{-1/2}$.}
\end{figure}

The resulting precision is shown in Figure \ref{fig:lab}. From scaling the noise without co-averaging any measurements, one would expect to achieve a precision of 206 ppm hr$^{-1/2}$ if the noise is uncorrelated in time. The mean precision over the 63 stars is very close to this limit at 230 ppm hr$^{-1/2}$, and we do not see any additional 1/$f$ noise when co-averaging up to 90 minutes of measurements.

\subsection{On-Sky Photometry}
\label{sec:sky}  
Finally, we tested the AP1121 camera on the sky using the MIT Wallace Observatory's 0.6 m telescope. We obtained $Y$-band differential photometry of GJ 3839 and TYC 2546-1103-1, with respective $J$ magnitudes\cite{2mass} of 8.443 and 9.678 and separated by 3.02'. The targets had an airmass of $\sim1.1$ during the observations, and the moon was below the horizon. Sky flats were taken in the evening twilight, and dark frames were obtained at the telescope as well. A focal reducer delivered a pixel scale of 0.873'' pix$^{-1}$ to the detector. Unfortunately, the seeing of $\sim$3.5'' and telescope tracking errors enlarged the point-spread function to a FWHM of $>$4 pixels. These tracking errors also prevent us from measuring the photometric precision on long time scales, but we can still measure the noise and compare it to our model.

The noise contributions to the sky testing are shown in Table \ref{tab:noise}. We have previously measured the effective aperture of the Wallace telescope to be 0.25 m \cite{sullivan13}, and the fluxes we measured are consistent with that value. The noise we measure on the sky is 20$\%$ higher than the noise we would expect from the star Poisson noise, dark current, sky background, and atmospheric scintillation.

\begin{table}[h]
\caption{Noise budget for a single exposure in laboratory and on-sky testing.} 
\label{tab:noise}
\begin{center}       
\begin{tabular}{|r|c|c|c|} 
\hline
\rule[-1ex]{0pt}{3.5ex} Noise Source & Lab Photometry & GJ 3839 & TYC 2546-1103-1 \\
\hline
\hline
\rule[-1ex]{0pt}{3.5ex} $J$ mag. & 8.45 (equivalent) & 8.443\cite{2mass} & 9.678\cite{2mass} \\
\hline
\rule[-1ex]{0pt}{3.5ex} Exposure Time [s] & 5 & 21 & 21 \\
\hline
\rule[-1ex]{0pt}{3.5ex} Number of pixels & 14 & 314 & 250 \\
\hline
\rule[-1ex]{0pt}{3.5ex} Read noise per pixel & $<$145 & 87 & 87 \\ 
\hline
\rule[-1ex]{0pt}{3.5ex} Counts from target [$10^5~e$-] & 2.4 & 9.1 & 3.5 \\
\hline
\hline
\rule[-1ex]{0pt}{3.5ex} Poisson noise [$e$-] & 490 & 910 & 590 \\
\hline
\rule[-1ex]{0pt}{3.5ex} Read noise [$e$-] & $<$540 & 1600 & 1420 \\
\hline
\rule[-1ex]{0pt}{3.5ex} Dark noise (200 $e$- pix$^{-1}$ s$^{-1}$) [$e$-] & 118 & 1150 & 1025 \\
\hline
\rule[-1ex]{0pt}{3.5ex} Sky noise [$e$-] & 0 & 410 & 370 \\
\hline
\rule[-1ex]{0pt}{3.5ex} Scintillation noise\cite{young67}$^,$\cite{young93}$^,$\cite{dravins97} [$e$-] & 0 & 900 & 360 \\
\hline
\hline
\rule[-1ex]{0pt}{3.5ex} Quadrature sum of noise sources & 740 & 2400 & 1920 \\
\hline
\rule[-1ex]{0pt}{3.5ex} Predicted Noise & 0.0031 & 0.0062 & (Differential) \\
\hline
\rule[-1ex]{0pt}{3.5ex} Measured Noise & 0.0026 & 0.0067 & 0.0071 \\
\hline

\end{tabular}
\end{center}
\end{table}

\clearpage

\section{SUMMARY AND DISCUSSION}
\begin{itemize}
\item{The dark current at -50$^{\circ}$C is below the $Y$ sky background for telescopes larger than $\left(1.0~\rm{m} \right) \left( \frac{\rm{Pixel~scale}}{1''} \right)^{-2}$. The higher sky background in the $J$ and $H$ bands reduces this minimum telescope size to 0.6 m and 0.3 m, respectively.}
\item{We mitigate read noise with up-the-ramp sampling. For exposures longer than 35 s, read noise is significant but should not dominate.}
\item{Nonlinearity is 5$\%$ across a dynamic range of 80k $e$- and is easily correctable with a second-order polynomial.}
\item{Persistence from the brightest observable targets should fall below the dark current within 10 s. For most targets, persistence should not be an issue. }
\item{Laboratory photometry is stable on long timescales at 230 ppm hr$^{-1/2}$, and we see no evidence of 1/$f$ noise after de-trending reference stars and centroid shifts.}
\item{On-sky photometry verifies our noise model to within 20$\%$.}
\end{itemize}
Although the AP1121 can deliver sky-limited performance on telescopes larger than $\sim$1.0 m (depending on the bandpass and pixel scale), we hope to further reduce the dark current and read noise to gain margin on this threshold. The dark current does not appear to depart from an exponential scaling with temperature, so an additional cooling stage should be advantageous. The read noise of the AP1121 is a more difficult problem to solve, but we are continuing to optimize the detector's configuration. 

If these modest improvements can be made to the performance of off-the-shelf InGaAs detectors in astronomical applications, they will become useful for 1-3 m telescopes, which could benefit from an economical source of focal planes for IR imaging. These detectors will probably never compete with high-performance HgCdTe detectors for spectrographs or space-based instruments, but the sky background relaxes the requirements for detector performance in ground-based imaging. Exoplanet transit searches from the ground, like the MEarth survey \cite{mearth}, could be extended to near-IR wavelengths to study late-M and L dwarfs for planets. In addition, a wide-field infrared telescope to follow-up gravitational wave triggers would complement the optical capabilities of the upcoming Zwicky Transient Factory and Large Synoptic Survey Telescope.

\acknowledgments     
 
We would like to thank Tim Brothers at MIT's Wallace Observatory for assisting with the observations described here. Hardware for this project was purchased with the MIT-Kavli Institute Development fund. BC's work was performed under contract with the California Institute of Technology funded by NASA through the Sagan Fellowship Program. The Natural Sciences and Engineering Research Council of Canada supports the research of BC. Lumbar support for RAS was provided through the Adam J. Burgasser Chair in Astrophysics.
\newpage

\bibliography{ingaas}   
\bibliographystyle{spiebib}   

\end{document}